\documentclass[aps,prb,twocolumn,groupedaddress,amsmath,amssymb]{revtex4}

\usepackage{textcomp}
\usepackage{color}
\usepackage{amsfonts}
\usepackage{bbold}
\usepackage{dsfont}
\usepackage{epsfig}
\usepackage{hyperref}

\newcommand{\arctanh}[1]{\operatorname{arctan}}

\begin{document}

\title{MgN: a new promising material for spintronic applications}

\author{A. Droghetti,  N. Baadji and S. Sanvito}
\affiliation{School of Physics and CRANN, Trinity College, Dublin 2, Ireland}

\date{\today}

\begin{abstract}
Density functional theory calculations demonstrate that rocksalt MgN is a magnetic material at the verge of half-metallicity, with an 
electronic structure robust against strong correlations and spin-orbit interaction. Furthermore the calculated heat of formation 
describes the compound as metastable and suggests that it can be fabricated by tuning the relative Mg and N abundance 
during growth. Intriguingly the equilibrium lattice constant is close to that of MgO, so that MgN is likely to form as an inclusion 
during the fabrication of N-doped MgO. We then speculate that the MgO/MgN system may represent a unique materials platform 
for magnetic tunnel junctions not incorporating any transition metals. 
\end{abstract}

\maketitle

%*********************************************************************
% Introduction
%*********************************************************************
The so called $d^0$ magnets are currently challenging our conventional understanding of magnetism. These are 
wide-gap semiconductors and oxide insulators displaying magnetic properties which cannot be ascribed to the 
presence of partially filled $d$ or $f$ shells\cite{coey}. Prototypical examples are diamagnetic materials in their 
well crystallized bulk form, which present magnetic features when grown as defective thin films. A similar 
phenomenology is found also in oxides doped with light elements such as N or C \cite{C_ZnO,C_ZnO2}. 
Unfortunately, despite the increasingly large number of reports of $d^0$ magnets, their typical experimental 
characterization is often limited to magnetometry with little information about the local electronic structure. The 
phenomenon thus remains rather obscure. It is characterized by a poor degree of reproducibility and 
still lacks of a convincing theoretical framework.

Most of the theoretical work to date is based on density functional theory (DFT) utilizing local approximations to the exchange 
and correlation functional (LDA or GGA). In most of the cases the LDA/GGA, probably correctly, associates the magnetic 
moment formation to spin-polarized holes residing on either cation vacancies or impurities. In addition the ground state
is always predicted ferromagnetic and rather robust. However LDA and GGA usually fail in describing the symmetry
details of the wave-function, in particular the polaronic distortion around the magnetic center. Therefore more advanced 
techniques, including strong electron correlation, appears as necessary \cite{me,Pic,saw,zunger}. When applied, these confirm the 
nature of the magnetic moment formation but almost always predict no long-range magnetic coupling and no ferromagnetism. 
MgO:N is a prototypical example\cite{me,Pic,me2,QMC}. Quantum Monte Carlo calculations for the Anderson-Haldane model
indicates the possibility of high temperature ferromagnetism\cite{QMC}. However DFT calculations using either the 
LDA+$U$ or the self-interaction correction (SIC) scheme return impurity levels deep in the gap, holes trapped by 
polaronic distortions and no room temperature ferromagnetism\cite{me,Pic,me2}.

Finally another important, but yet scarcely studied, class of $d^0$ magnets is represented by zincblende (ZB) II-V or II-IV 
compounds such as CaP, CaAs, CaSb \cite{Ku,Bo,Si} or CaC, SrC and BaC\cite{Sandra}. These are weakly covalent
solids with either one or two holes per formula unit, and they are generally predicted to be half-metallic. Intriguingly, 
although at least for the II-V's the thermodynamical stable phase has the stoichiometric Zn$_3$P$_2$-type structure\cite{Bo2}, 
their rocksalt (RS) phase has usually a lower energy than the ZB and it is predicted metastable. 

In this work we investigate the magnetic ground state of RS MgN, which was already predicted as half-metal in the 
ZB phase \cite{Si}. This is an extremely interesting material because of its structural similarity to the parental MgO,
widely used as tunnel barrier in magnetic tunnel junctions~\cite{Parkin}. We will demonstrate that MgN is
a half-metal with negligible spin-orbit interaction and negative formation enthalpy. This means that, although it is
not the thermodynamical stable phase, it can be grown as a metastable compound in the form of either thin films
or as nanoclusters inside MgO matrices. 

Our calculations are performed with a development version of the {\sc siesta} code\cite{19}, which includes the 
atomic self interaction correction (ASIC)\cite{20} and the LDA+$U$\cite{Tom} schemes. The core electrons are treated 
with norm-conserving Troullier-Martin pseudopotentials and the valence charge density is expanded over a numerical 
orbital basis set, including double-$\zeta$ and polarized functions \cite{19}. The real space grid has an equivalent
cutoff larger than 600~Ry, and we sample a minimum of $20\times20\times20$ $k$-points in the Brillouin zone. The 
atomic coordinates are relaxed by conjugate gradient until the forces are smaller than 0.01~eV/\AA. For the evaluation 
of the spin-orbit coupling we have performed full-potential linear augmented plane-waves (FLAPW) calculations 
\cite{nadj15} with the scheme implemented in {\sc fleur} code \cite{nadj17}. We use a $12\times12\times12$ 
$k$-point mesh and a plane wave expansion with $k_\mathrm{max}=4$ a.u. A muffin-tin sphere of a radius $2.02$ 
Bohr was taken for both Mg and N with $\ell_\mathrm{max}$=8 as cutoff in the  $\ell$-expansion of the muffin orbitals. 
Spin-orbit  is included by second variation method\cite{ soc}. 

We start our analysis by briefly comparing the ground state of the ZB and RS structures. In table~\ref{Tab1} we report
the equilibrium lattice constant, $a$, the magnetic moment per formula unit, $\mu$, and the magnetic moment 
formation energy, $\Delta E_\mathrm{M}$, for the two structures. $\Delta E_\mathrm{M}$ is defined as the energy
difference per formula unit between the spin-polarized and the non-spin-polarized DFT solution and it is positive
when the magnetic state has a lower energy. Importantly $\Delta E_\mathrm{M}$ is always substantially larger than 
zero, demonstrating that the magnetism is robust for both the structures. We also note that our results for the ZB phase 
are in excellent agreement with previous calculations\cite{Si}, and that for the RS phase $\Delta E_\mathrm{M}$ is 
slightly smaller than that calculated for the same phase of SrN and CaN\cite{Bo2}, although it is still well above room 
temperature.
 \begin{table}[h]
\begin{tabular}{lccccccc} \hline\hline
Structure & \multicolumn{3}{c}{LDA} &  \multicolumn{3}{c}{GGA} \\ \hline
& $a$ & $\mu$ & $\Delta E_\mathrm{M}$ & $a$ & $\mu$ & $\Delta E_\mathrm{M}$ \\
 &  &  &  &  &  &  \\
ZB & 4.73 & 1.0 & 157 & 4.83 & 1.0 & 180 \\
RS & 4.34 & 1.0 & 60   & 4.44 & 1.0 & 95 \\ 
\hline\hline
\end{tabular}\caption{\label{Tab1}Equilibrium lattice constant $a$ (in \AA), magnetic moment per formula unit $\mu$ 
(in $\mu_\mathrm{B}$) and $\Delta E_\mathrm{M}$ (in meV) for ZB and RS MgN.}
\end{table}

\begin{figure}[ht]\centering
\includegraphics[scale=0.2,clip=true]{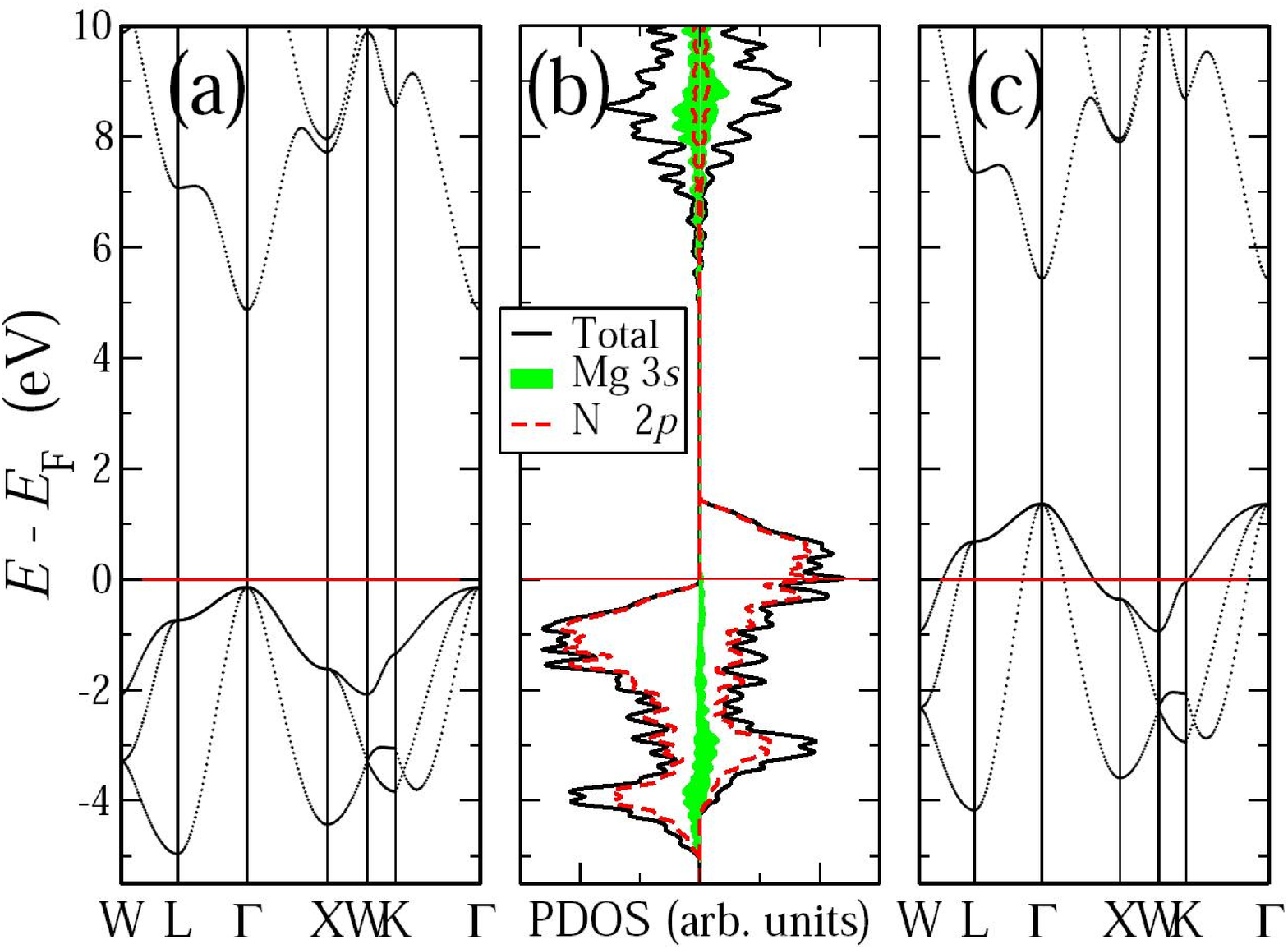}
\caption{GGA band structure for RS MgN calculated at the equilibrium lattice constant of 4.44~\AA: panel (a) majority spin, panel (c) minority.
The horizontal red line denotes the position of Fermi level. In panel (b) we report the associated total density of states (black solid line) 
and density of states projected over the N-$2p$ (red dashed line) and Mg-$3s$ (green shadowed area)}\label{MgN_bands}
\end{figure}

Next we discuss in details the electronic structure of RS MgN. The GGA bandstructure calculated at the equilibrium lattice 
constant of 4.44~\AA\ and displayed in Fig.~\ref{MgN_bands}, clearly reveals the half-metallic character of MgN with 
a direct gap of 5~eV at the $\Gamma$ point in the minority spin-channel. This is different from the bandstructure for 
RS CaN where the gap is indirect with the conduction band minimum located at X \cite{Bo2}.
The differences between CaN and MgN can be understood by looking at the orbital projected density of states (PDOS) presented in Fig.~\ref{MgN_bands}(b).
In the case of MgN the N 3$d$ levels are unbound and they only weakly mix with the Mg 3$s$, which form the conduction band. In contrast the empty
Ca 3$d$ levels provide the dominant contribution to the conduction band of CaN, shifting the band minimum to the X point. Furthermore their 
hybridization with the N 2$p$ orbitals forming the valence band is rather strong and result in the CaN valence band being considerably more
flat than that of MgN. In summary the main difference between RS MgN and CaN is in the different role played by the cation
3$d$ orbitals. Interestingly it was suggested\cite{Ku} that the contribution of the Ca-$d$ level in Ca pnictides is fundamental to promote the 
half metallicity. Later such a claimed was challenged by the statement that the magnetic moment formation in II-V compounds is 
driven by the strong atomic character of the anions\cite{Bo2}. MgN demonstrates that both of the arguments are partially correct. Indeed the 
ionicity in SrN, CaN and MgN is able to localize the hole on N leading to the formation of the magnetic moment. However the hybridization 
with the $d$ shell of the cation increases the flatness of the top of valence band causing an enhancement of $\Delta E_\mathrm{M}$.

\begin{figure}[ht]\centering
\includegraphics[scale=0.2,clip=true]{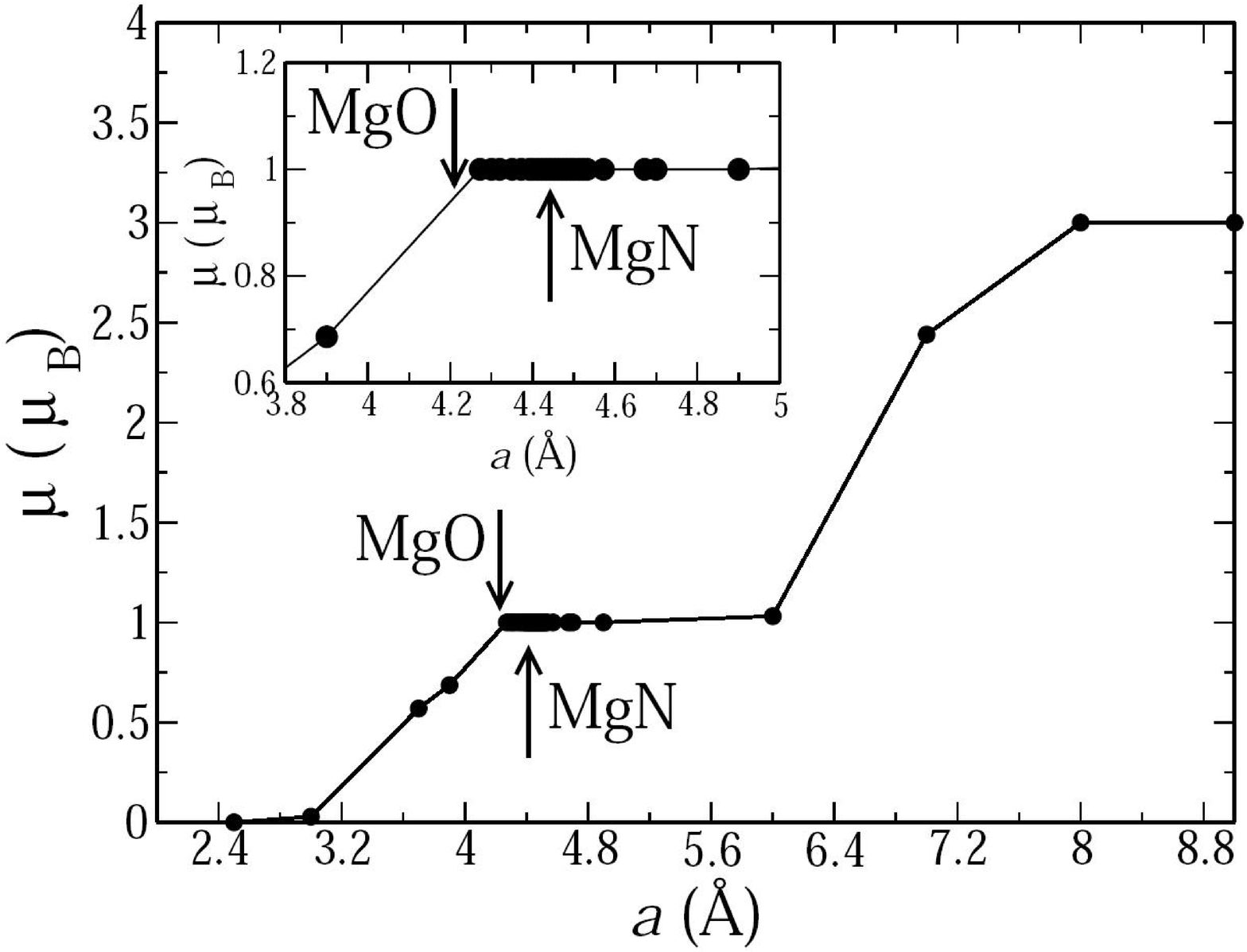}
\caption{Magnetic moment $\mu$ per formula unit as a function of the lattice constant for RS MgN as calculated with GGA.
The inset shows a zoom at around the GGA equilibrium value of 4.44~\AA. The arrows indicate the experimental MgO and the
GGA-calculated MgN lattice constants.}
\label{MgN_spin_lattice}
\end{figure}

We then investigate the robustness of the half-metallic ground state with respect to compression, spin-fuctuations, choice of exchange and
correlation functional and spin-orbit interaction. In figure~\ref{MgN_spin_lattice} we report the magnetic moment as a function of the
lattice constant as calculated with GGA (the LDA results are rather similar). Importantly we note that there is a broad range of lattice constants 
for which MgN preserves the half-metal moment of 1~$\mu_\mathrm{B}$. For $a$ smaller than 4.3~\AA\ the moment 
becomes fractional and the material turns into a standard ferromagnet, until it finally becomes non-spin-polarized for a massive compressive 
pressure ($a<3.0$~\AA). Interestingly at the MgO lattice constant (4.21~\AA) the magnetic state of MgN is still extremely 
close to that of a half-metal. This is a tantalizing feature suggesting that, should MgN be made in MgO, it will be an efficient spin
polarizer/analyzer in transition metal free tunnel junctions. Finally note that for a largely expanded lattice parameter the moment reaches the 
value of 3~$\mu_\mathrm{B}$ expected from isolated N ions. 

Spin excitations and the possible effects arising from strong correlations are analyzed next. In particular we calculate the total energy difference 
per formula unit between the ferromagnetic and either the type-I ($ \Delta E_{\uparrow\uparrow-\uparrow\downarrow}^\mathrm{I}$) and 
type-II ($ \Delta E_{\uparrow\uparrow-\uparrow\downarrow}^\mathrm{II}$) antiferromagnetic state. These are both characterized by a planar 
ferromagnetic alignment with antiparallel orientation between planes, either along the (001) (type-I) or (111) (type-II) direction. At the equilibrium
lattice constant we find $\Delta E_{\uparrow\uparrow-\uparrow\downarrow}^\mathrm{I}=52$~meV 
($ \Delta E_{\uparrow\uparrow-\uparrow\downarrow}^\mathrm{I}=67$~meV) and 
for $ \Delta E_{\uparrow\uparrow-\uparrow\downarrow}^\mathrm{II}=53$~meV 
($ \Delta E_{\uparrow\uparrow-\uparrow\downarrow}^\mathrm{II}=68$~meV) for LDA (GGA). These are rather similar to $\Delta E_\mathrm{M}$
indicating that in MgN Stoner and spin-wave excitations compete. 

It is then interesting to investigate how the moment formation and the energetic of the spin-excitations change when electron correlation is
added to the electronic structure. This is investigated with the LDA+$U$ scheme applied to the N-$2p$ shell as a function of $U$ 
($J$ is kept at zero). Our results are reported in table~\ref{Tab2}. As expected the addition of on-site correlation to the N-$2p$ shell downshifts
its center of mass effectively opening the band-gap in both the spin-bands. This is associated to a considerable strengthening 
of the ferromagnetic interaction, with both 
$\Delta E_{\uparrow\uparrow-\uparrow\downarrow}^\mathrm{I}$ and $\Delta E_{\uparrow\uparrow-\uparrow\downarrow}^\mathrm{II}$
increasing monotonically with $U$. In the table we also report the exchange split, $\Delta E_\mathrm{ex}$, which is the energy difference
between the top of the majority and minority valence bands for the ferromagnetic ground state at its equilibrium lattice constant. 
We note that also $\Delta E_\mathrm{ex}$
increases with $U$ with an almost perfectly linear dependence. Finally we find that the critical lattice constant for the half-metallicity gets 
progressively reduced as $U$ gets larger, while the equilibrium lattice parameter seems to be only marginally affected by $U$. 

The promotion of the magnetic stability in RS MgN with increasing $U$ can be understood by using the DFT version of the standard Stoner 
argument applied to the LDA+$U$ functional\cite{Pet}. In fact, in the limit of uniform occupancy of the strongly correlated shell (N-2$p$ in
our case), one expects the Stoner parameter to increase linearly with $U$. This may, of course, be compensated by a reduction of the 
density of states at the Fermi energy ($E_\mathrm{F}$), resulting in an overall reduction of the magnetic interaction. In the case of RS MgN, 
however, both the band-width and the curvature of the bands near $E_\mathrm{F}$ change little with $U$, since such a valence band has a
pure N-2$p$ character. As a consequence we find an almost perfect linear relation between the exchange split $\Delta E_\mathrm{ex}$
and $U$, which confirms the Stoner nature of the ferromagnetism in MgN. This leads us to conclude that the addition 
of strong correlations through an on-site Coulomb repulsion enhances the stability of the half-metal character of RS MgN. 

 \begin{table}[h]
\begin{tabular}{clccccc} \hline\hline
$U$ & $a_\mathrm{c}$ & $E_\mathrm{gap}$  & $\Delta E_\mathrm{ex}$ & $\Delta E_{\uparrow\uparrow-\uparrow\downarrow}^\mathrm{I}$  &
$\Delta E_{\uparrow\uparrow-\uparrow\downarrow}^\mathrm{II}$ \\ 
(eV) & (\AA) & (eV) & (eV) & (meV)  & (meV) \\ \hline\hline
GGA & 4.3 & 5.0 & 1.51 & 52 & 53 \\
2 & 3.8 & 6.19 & 2.15 & 95 & 96 \\
3 & 3.6 & 6.40 & 2.48 & 172 & 156  \\ 
4 & 3.5 & 6.75 & 2.94 & 267 & 251 \\ 
5 & 3.4 & 7.16 & 3.30& 344 & 338  \\ 
6 & 3.4 & 7.40 & 3.77 & 395 & 392 \\ 
7 & 3.3 & 8.09 & 4.23 & 443 & 428\\ 
8 & 3.2 & 8.17 & 4.68 & 472 & 463 \\
ASIC & 3.6 & 7.51& 4.24 & 184 & 175  \\
\hline\hline
\end{tabular}\caption{\label{Tab2}Critical lattice constant for the half-metallicity, $a_\mathrm{c}$, minority spin bandgap, $E_\mathrm{gap}$,
exchange split $\Delta E_\mathrm{ex}$, and ferromagnetic to antiferromagnetic total energy difference
$\Delta E_{\uparrow\uparrow-\uparrow\downarrow}^\mathrm{I}$ and $\Delta E_{\uparrow\uparrow-\uparrow\downarrow}^\mathrm{II}$,
as a function of the on-site Coulomb interaction $U$.}
\end{table}
In table~\ref{Tab2} we also report results obtained with the ASIC method, which was proved to yield an accurate MgO bandgap and the
proper polaronic distortion in diluted $d^0$ magnets\cite{me}. The ASIC returns both a critical lattice constant for the half-metallicity and
magnetic energies similar to those calculated with LDA+$U$ and $U\sim 3$~eV. However the bandgap and $\Delta_\mathrm{ex}$
are considerably larger and compatible with $U=6$~eV. Although it is difficult to make any strong quantitative statement, because the 
mixed success of ASIC in predicting exchange constants\cite{Akin}, the result further confirms that the half-metallic state of MgN is strong 
with respect to strong electron correlation.  

Finally we investigate the effect of the spin-orbit coupling on the electronic structure of MgN. In general spin-orbit interaction always destroys
the half-metallicity of a magnet by mixing the two spin channels. Our calculations however show that the spin-orbit coupling constant between 
the spin and the orbital angular momentum is only of the order of $10$~meV. Furthermore, the MgN highly symmetric cubic enviroment 
quenches the angular magnetic moment, which is only $7\times 10^{-5}$~$\mu_\mathrm{B}$ for Mg and $9.8 \times 10^{-4}$~$\mu_\mathrm{B}$ 
for N. It is then not surprising that the spin polarization remains 100~\% even after the spin-orbit interaction
is accounted for, i.e. the half-metallicity is preserved. 

\begin{figure}[ht]\centering
\includegraphics[scale=0.2,clip=true]{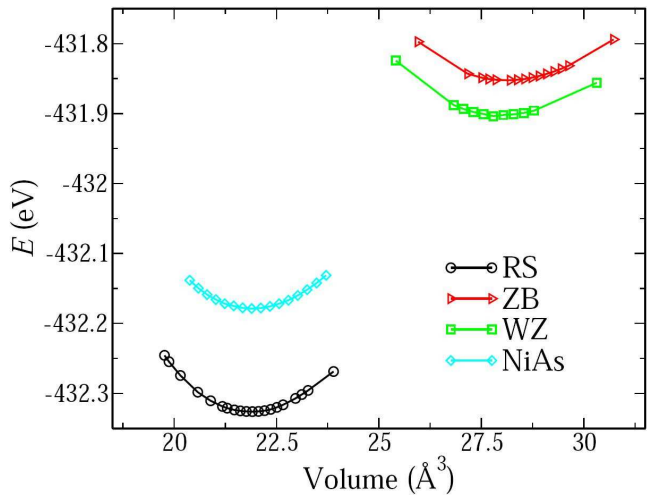}
\caption{(Color on line) Total energy as a function of the volume per formula unit for MgN with wurtzite (WZ), 
zincblende (ZB), NiAs and rock-salt (RS) crystal structure.}\label{MgN_volume}
\end{figure}
After having discussed the electronic properties of MgN we now turn our attention to analyzing the structural stability of the RS structure.
First we compare the total energy per cell of volume of four different crystal structures, likely candidates for MgN,
namely wurzite, ZB, NiAs-type (space group $P6_3/mmc$) and RS. As shown in figure~\ref{MgN_volume} 
the RS structure presents both the lower total energy and smaller equilibrium volume of all the crystals investigated, indicating 
that it is the most thermodynamic favorable. This however still does not mean that MgN can be made. In fact one has to compare
the stability of RS MgN with that of its stechiometric stable form Mg$_3$N$_2$, i.e. one has to demonstrate that the heat of formation
for RS MgN, $\Delta H_\mathrm{MgN}$, is comparable to that of Mg$_3$N$_2$, $\Delta H_\mathrm{Mg_3N_2}$. 

Mg$_3$N$_2$ is a non-magnetic semiconductor with an experimental band gap of $2.8$~eV\cite{mg3n22} and with the tetragonal 
bixbyite crystal structure. The experimental lattice constant is $9.953$~\AA\cite{mg3n2} while our calculations return values of 
$9.85$~\AA\ (LDA) and $10.01$~\AA\ (GGA). These are both within 1\% of the experimental reported values. The heats of formation 
can be written as
\begin{eqnarray}
\Delta H_\mathrm{Mg_3N_2}=1/5\left[1/8\:E_\mathrm{Mg_3N_2}-\left(3/2\:E_\mathrm{Mg} +E_\mathrm{N_2}\right)\right],\\
 \Delta H_\mathrm{MgN}=1/2\left[E_\mathrm{MgN}-1/2\left(E_\mathrm{Mg} +E_\mathrm{N_2}\right)\right],
\end{eqnarray}
where $E$ refers to the total energy and Mg is assumed in its $hcp$ metallic phase. We
calculate $\Delta H_\mathrm{Mg_3N_2}=-1.44$~eV/atom ($\Delta H_\mathrm{Mg_3N_2}=-1.16$~eV/atom) and 
$\Delta H_\mathrm{MgN}=-0.91$~eV/atom ($\Delta H_\mathrm{MgN}=-0.59$~eV/atom) with LDA (GGA). Although, as expected 
Mg$_3$N$_2$ has a lower heat of formation and forms at equilibrium, RS MgN also has a negative $\Delta H$. 
This means that it is a thermodynamic stable compound, and hence that RS MgN is metastable. We then
speculate that the relative abundance of N and Mg during the growth process, in addition to the choice of substrate, can drive the 
formation of either Mg$_3$N$_2$ or MgN. Intriguingly MgO appears as an ideal substrate for the growth of MgN but metals with bcc lattice 
structure are equally tantalizing. Among them we mention V, Mo and W.

In conclusion we investigated the electronic and structural properties of rocksalt MgN. This was found to be at the verge of half-metallicity in
LDA and GGA and completely half-metallic as soon as some electron correlation is included. The half-metallicity is then robust with compression
and spin-orbit interaction. Furthermore we have then demonstrated that the RS phase is the most stable among the various possible MgN 
phases. Its heat of formation is negative, but smaller than that of Mg$_3$N$_2$, suggesting that MgN can be possibly made by tuning the 
relative Mg and N abundances during growth and the substrate crystal structure.

This work is sponsored by Science Foundation of Ireland (07/RFP/MASF238) and by the EU 6$^\mathrm{th}$ Framework
(SpiDME project). Computational resources have been provided by the HEA IITAC project managed by the Trinity Center 
for High Performance Computing and by ICHEC.

\end{document}